\documentclass[aps,twocolumn,prb,superscriptaddress,groupaddress,showpacs,preprintnumbers,amsmath,amssymb]{revtex4}
\usepackage{graphicx, bm}
\usepackage{float}
\usepackage{latexsym}
\usepackage{amsmath}
\usepackage{graphics}
\usepackage{amssymb}
\usepackage{layout}
\usepackage{verbatim}
\usepackage{amsfonts,epsfig}
\usepackage{color}
\usepackage{setspace}
\newcommand{\beqnar}{\begin{eqnarray}}
\newcommand{\eeqnar}{\end{eqnarray}}

\newcommand{\me}{\mathrm{e}}

\newcommand{\bk}{{\bf k }}

\newcommand{\br}{{\bf r }}

\newcommand{\bv}{{\bf v }}

\newcommand{\bG}{{\bf G}}

\newcommand{\beq}{\begin{equation}}
\newcommand{\eeq}{\end{equation}}

\begin{document}
\title{Anisotropic surface transport in topological insulators in proximity to a helical spin density wave}
\author{Qiuzi Li}
\affiliation{Condensed Matter Theory Center and Joint Quantum Institute, Department of Physics, University of Maryland, College Park, Maryland, 20742-4111, USA}
\author{Parag Ghosh}
\affiliation{Department of Physics and Astronomy, George Mason University, Fairfax, Virginia, 22030, USA and \\ 100 Bureau Drive, Stop 8410, NIST, Gaithersburg, Maryland, 20899-8410, USA}
\author{Jay D. Sau}
\affiliation{Condensed Matter Theory Center and Joint Quantum Institute, Department of Physics, University of Maryland, College Park, Maryland, 20742-4111, USA}
\author{Sumanta Tewari}
\affiliation{Department of Physics and Astronomy, Clemson University, Clemson, South Carolina, 29634, USA}
\author{S. Das Sarma}
\affiliation{Condensed Matter Theory Center and Joint Quantum Institute, Department of Physics, University of Maryland, College Park, Maryland, 20742-4111, USA}
\date{\today}
\begin{abstract}
We study the effects of spatially localized breakdown of time reversal symmetry on the surface of a topological insulator (TI) due to proximity to a helical spin density wave (HSDW). The HSDW acts like an externally applied one-dimensional periodic(magnetic) potential for the spins on the surface of the TI, rendering the Dirac cone on the TI surface highly anisotropic. The decrease of group velocity along the direction $\hat{x}$ of the applied spin potential is twice as much as that perpendicular to $\hat{x}$. At the Brillouin zone boundaries (BZB) it also gives rise to new semi-Dirac points which have linear dispersion along $\hat{x}$ but quadratic dispersion perpendicular to $\hat{x}$. The group velocity of electrons at these new semi-Dirac points is also shown to be highly anisotropic. Experiments using TI systems on multiferroic substrates should realize our predictions. We further discuss the effects of other forms of spin density wave on the surface transport property of topological insulator.
\end{abstract}
\maketitle

\section{Introduction}

In the last few years there has been a growing interest in topological insulators (TI), which are materials that are insulators in the bulk but conduct two-dimensionally on the surfaces \cite{qi_prb_2008, hsieh_science_2009, zhang_nature_2009, roy_prb_2009, chen_science_2009, kane_rmp}. The non-trivial topology of the wavefunction of such TI has been
predicted to show metallic surface conductivity that is topologically
 protected against weak disorder and interaction effects.
 One of the mechanisms by which the surface modes can be disrupted
 is by breaking the time reversal invariance upon applying
 a magnetic field. There have been several studies on the effects
 of a magnetic field either applied directly perpendicular to the
surface of a TI \cite{tse_prl_2010} or by the proximity effect of
 a ferromagnet in a heterostructure
\cite{mondal_prl_2010, garate_prl_2010}. On the other hand,
 the effects of an antiferromagnet or a spin density wave on the
surface states of the TI is a topic which is much less well-explored. We carry out such an analysis of the effect of a spin density wave on TI transport in this work.

Although an antiferromagnet, in the proximity to a TI, does not affect the global time reversal symmetry (TRS) in the latter, the staggered nature of the spins in the antiferromagnet breaks the TRS locally. In this work, we are interested in how this local TRS breaking affects the surface states of the TI. However, the local TRS breaking occurs at a length scale which is of the order of the lattice spacing in the antiferromagnet (for e.g., $8.85 \AA$ for MnO). Therefore the effects of local TRS breaking occur at very large momenta, which makes it difficult to observe them in experiments. In order to redress this difficulty, we propose to study the effects of local TRS breaking by replacing the antiferromagnet with a multiferroic material\cite{cheong_nature_2007} like orthorhombic RMnO$_3$ (R being a rare earth element like Tb or Dy) or the family of Fe$_{1-x}$Co$_x$Si with cubic but noncentrosymmetric structure, which shows a helical spin density wave (HSDW) order with relatively long periods ($>30$nm) \cite{uchida_science_2006}. The HSDW will couple to the  Dirac fermions on the surface of the topological insulator due to the proximity effect by the above two kinds of materials directly deposited on the top. The situation is similar to that of an externally applied charge potential on graphene \cite{Louie_NPH08}. However, the contrast between the two situations is that the 3D topological insulator has spin polarized 2D Dirac fermions on its surface \cite{Hasan_NPH09}, whereas the Dirac fermions in graphene carry pseudo-spin. Nevertheless, in analogy with the charge potential applied to graphene, a spin potential can be used for TI to manipulate the surface Dirac fermions.

It may be appropriate to ask about the motivation of our theoretical work where we are proposing, using concrete theoretical calculations, that experiments be carried out on a multiferroic-TI sandwich structure for the observation of the topologically protected surface transport in the TI surface states induced by the HSDW associated with the multiferroic layer.  This approach sounds somewhat indirect, and indeed it is although the effect of the HSDW-induced local TRS-breaking on TI surface transport properties is intrinsically interesting in its own right as we establish in this work.

Our main motivation for this work, however, arises from the fact
that so far there has been little direct experimental signature of
the topologically-protected surface transport in existing TI
materials in spite of a great deal of research activity
\cite{Ren_PRB10, Qu_Science10, Butch_PRB10, Analytis_nph_2010,
Culcer_PRB10, Checkelsky_arXiv10, Taskin_PRB10, Akinori_PRB10,
Eto_PRB10}  aimed precisely toward the direct and unambiguous
observation of 2D surface TI transport.  All the convincing
experimental evidence for the existence of the
topologically-protected surface 2D TI states is currently based on
the verification of the proposed band structure through STM/STS or
ARPES type spectroscopic measurements.  The problem in the direct
observation of the surface 2D TI transport is that the currently
existing TI materials are not good bulk insulators, and the bulk
conduction is \textit{always} much stronger than the surface
conduction, making it impossible to observe the surface 2D states
unambiguously in transport measurements.  This problem of
substantial bulk transport arises from the fact that all existing TI
materials, instead of being a bulk band insulator as band structure
calculations predict, turn out to be intrinsically doped in the bulk
due to defects and vacancies, giving rise to a large bulk conduction
channel which competes directly with the surface 2D transport.   For
example, the two most recent (and also most compelling) transport
measurements \cite{Ren_PRB10, Qu_Science10} see very small putative
surface 2D magneto-resistance oscillations with a temperature
dependent bulk resistivity which does not behave like a standard
band insulator at all.   In particular, the bulk resistivity in
these samples \cite{Ren_PRB10, Qu_Science10} is of the order of
$\Omega\cdot$cm or less whereas an insulator typically has a bulk
resistivity which is 10--12 orders of magnitude larger.  Thus, the
TI systems, even in these most compelling measurements, show
dominant bulk conduction with less than 1\% of the net conduction
being inferred (through the indirect fitting of the
magneto-resistance data using many free parameters and ad hoc
multichannel conduction models) at best to be arising from the 2D
surface states. Another recent experiment \cite{Butch_PRB10}
concludes that no 2D surface conduction can be discerned at all in
the TI transport data because of the dominant bulk conduction.
Another recent experiment \cite{Analytis_nph_2010} goes to the
extreme of using a pulsed external magnetic field as high as 55T in
order to investigate the surface 2D transport, again emphasizing
small observed features in the high-field resistivity as the
possible manifestation of the expected 2D TI surface transport. Even
the two very recent transport measurements \cite{Ren_PRB10,
Qu_Science10} purportedly claiming the manifestation of 2D surface
TI transport can only observe small 2D features in the derivative
spectrum of the resistivity with respect to the applied magnetic
field.  This is a most unsatisfactory state of affairs in sharp
contrast to the transport properties of well-established 2D quantum
systems, e.g. graphene, where 2D transport behavior
\cite{DasSarma_arXiv10} without any bulk conduction problem
whatsoever manifests itself in every possible transport measurement
in a decisive manner, and one does not have to look for small
features in the derivative spectrum.  The current situation in TI
physics is thus extremely problematic with all spectroscopic
measurements providing reasonable verification of the expected
spin-resolved 2D Dirac cone band structure on the TI surface whereas
the fact that the bulk is not an insulator is making it essentially
impossible to see the expected 2D protected surface transport.  We
point out in this context that the theoretical details of how the
surface 2D TI transport would behave, had it not been contaminated
by the unintentional bulk conduction problem plaguing the existing
TI materials, are reasonably well-known in the literature
\cite{Culcer_PRB10}.

Of course, further materials development in the existing TI systems leading to the effective suppression of the unintentional bulk doping  or the discovery of completely new classes of TI materials where the bulk is a true insulator  could solve this problem instantaneously, but until that happens, any idea which points toward the observation of surface TI transport should be welcome.  It is somewhat of an embarrassment that in spite of the huge activity in TI research, a true Topological Insulator does not yet exist since the existing systems are not true bulk insulators due to the invariable presence of unintentional bulk doping.  Our proposal in the current work should be seen in this light. We provide a method to directly isolate surface transport features in TI systems in proximity to a helical spin density wave as produced, for example, by a multiferroic material.  Since bulk conduction should be relatively immune to the presence of the HSDW, whereas the surface TI conduction should be affected qualitatively as we show in this work,  we believe that our predictions could help provide the unambiguous observation of the surface 2D TI transport properties.  In addition, the coupling between the HSDW and the 2D TI states leads to nontrivial novel physics (e.g. the semi-Dirac points to be discussed below) which is intrinsically interesting in its own right.

We find that the surface states of the TI are significantly modified in the presence of the HSDW. The HSDW acts like an externally applied superlattice potential on the TI surface resulting in striking anisotropy of the Dirac cones and group velocity of the surface states. In particular, the group velocity along the direction of the superlattice is monotonically suppressed as a function of the lattice potential strength and its period. For the type of HSDW considered
in Sec. \ref{sec:phsdw}, which we call proper HSDW, \cite{kataoka_JPSJ_1981} (see below), we find that novel semi-Dirac points, whose low-energy characteristics are intermediate between Dirac (massless) and zero-gap (massive) semiconductors \cite{Pickett_PRL_2009,Pardo_PRL_2009}, show up on the BZB. In Sec. \ref{sec:other}, we discuss how the surface transport properties of the topological insulator changes when the forms of the spin density wave is changed. Section \ref{sec:summary} contains a summary and conclusions.

\section{The proper helical spin density wave}
\label{sec:phsdw}
\subsection{Dirac cone on the BZ center}
We begin by writing down the effective Hamiltonian for the low-energy quasiparticles on the surface of a topological insulator as,
\begin{equation}
H_0(\bk) = \hbar v_F (k_x \sigma_x+ k_y \sigma_y)
\label{ham}
\end{equation}
where $v_F \simeq 6.2\times10^5$m/s is the Fermi velocity of Dirac fermions in Bi$_2$Se$_3$ \cite{SCZh_NPH09} and $\sigma_i$ are the Pauli matrices. The Hamiltonian in Eq. (\ref{ham}) is characterized by an energy spectrum $\varepsilon_{s,k}=s\hbar v_F k$
where $s = \pm 1$ is the band index, and eigenstates given by
\begin{equation}
\left<\br|s,\bk\right>=\frac{1}{\sqrt{2}}
e^{i\bk\cdot\br}\left(
\begin{array}{c}
1\\
se^{i\theta_{\bk}}
\end{array}
\right)\ ,
\end{equation}
where $\theta_\bk$ is the angle of vector $\bk$ with respect to the $\hat k_x$ direction.

\begin{figure}\includegraphics[width=0.9\columnwidth]{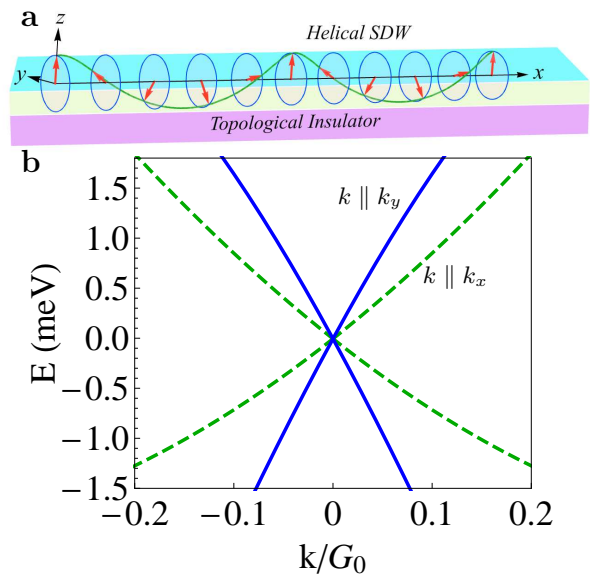}
 \caption{(Color online)
  (a) Schematic diagram of the system. The HSDW is on top of the topological insulator with the periodicity $L$. (b)
   Energy of charge carrier dispersion versus the wavevector $k$ with $k$ measured near the BZ center.
  Dashed (green) and solid (blue) lines show the
dispersion along the $\hat{x}$ direction  and
 along the $\hat{y}$
direction, respectively with $U_0=50$meV and $L=90$nm ($E(0,0)$ was shifted to zero). }
  \label{fig:S}
\end{figure}

In the presence of the HSDW on top of a TI, depicted schematically in Fig. \ref{fig:S}, the Hamiltonian (\ref{ham}) is modified \cite{Nagaosa_PRB10} to: $H= H_0(\bk) + U(x)$, where the potential $U(x)$ for the HSDW can be written as:
\begin{equation}
\begin{array}{l l }
U(x)= U_y \sigma_y \cos(\dfrac{2\pi}{L}x)+ U_z \sigma_z \sin(\dfrac{2\pi}{L}x),
\end{array}
\end{equation}
where $L$ is the spatial period of the potential, and $U_{y,z}$ are the amplitudes of the HSDW\cite{kataoka_JPSJ_1981}. For the case $U_z< U_y$, we find no gapless states on the BZB.
 On the other hand, for $U_z>U_y$, two Dirac points centered at $(\frac{\pi}{L},\pm K_y)$ emerge on the BZB, which will be discussed in Sec. \ref{sec:other}.  For the symmetric HSDW such that
 $U_z = U_y = U_0$, which we shall call proper HSDW, we find a single
 semi-Dirac point at $(k_x, k_y)=(\pm\dfrac{\pi}{L},0)$
 (Fig. \ref{fig:delta}a). In this section, we shall focus on the case of a
 proper HSDW on TI.

 In our calculation, we use the value of induced exchange field due to the magnetic proximity effect being $5 \sim 50$ meV\cite{Haugen_PRB08, Chakhalian_nph_2006}, which has been taken to be a reasonable value\cite{Nagaosa_PRB10}. The exchange-induced potential $U(x)$ from the HSDW creates a periodic
potential on the surface of the TI. The electronic eigenstates of such a
Hamiltonian can be written using Bloch's theorem as
\begin{equation}
\psi_{n\bm k}(\bm r)=\sum_{\bm G}e^{i (\bm k+\bm G)\cdot\bm r}c(n,\bm k+\bm G)
\end{equation}
where $\bG = \dfrac{2m\pi}{L}\hat{x}$ ($m\in$ integer) is the reciprocal lattice vector, $\psi_{n\bm k}(\bm r)$ and $c(n,\bm k+\bm G)$ are 2-spinor functions
and $n$ is the band-index. Here $k_x$ is only limited to be in the first
Brillouin Zone (FBZ), i.e. $k_x\in [-\dfrac{\pi}{L},\dfrac{\pi}{L}]$.
The band eigenstates $\psi_{n\bm k}(\bm r)$ and the corresponding
eigen-energies $E_n(\bm k)$ are obtained as eigenvalues and
eigenvectors of the Bloch equation
\begin{align}
&[\hbar v_F ((k_x+G)\sigma_x+ k_y \sigma_y)-E_{n\bk}]c(n,\bm k+\bm G)\nonumber\\
&+(U_y\sigma_y+i U_z\sigma_z)c(n,\bm k+\bm G-\dfrac{2\pi}{L})/2\nonumber\\
&+(U_y\sigma_y-i U_z\sigma_z)c(n,\bm k+\bm G+\dfrac{2\pi}{L})/2=0\label{eq:linear}.
\end{align}

While the above Bloch equation may be solved by numerical diagonalization to obtain numerical eigenvalues and eigenvectors, more insight
can be obtained into the solutions around high symmetry points in the FBZ
 at $\bm k=0$ and at the BZB by using perturbation theory. The numerical
results of diagonalizing Eq.~\ref{eq:linear} are shown in Fig. 2, Fig. 3 and Fig. 4.
The total Hamiltonian $H_0+U$ of the TI+HSDW system is invariant under a composite symmetry $S=\Theta T$ where $\Theta$ is the time-reversal operator and $T$ is the operator corresponding to translation by $L/2$. Since, the operator $S$, similar to the time-reversal symmetry operator $\Theta$, is
both anti-unitary and satisfies $S^2=-1$, the proximity to the HSDW does not open a gap
at $\bm k=0$.
 In other words, the HSDW preserves the global time-reversal symmetry, it does not open up a
gap near $\bm k=0$ on the TI surface.

\begin{figure}\includegraphics[width=1.0\columnwidth]{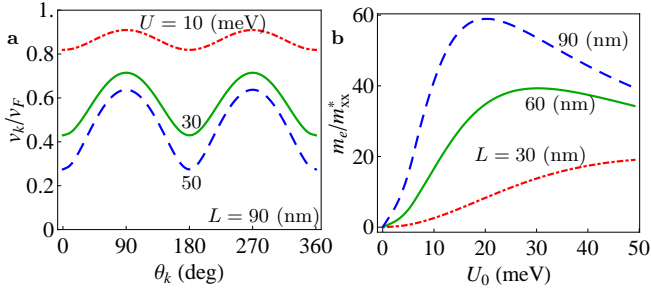}
  \caption{(Color online)
  (a) The group velocity $v_{\hat{k}}$ (measured from the Dirac point)
  of charge carriers on the surface of the TI in units of the Fermi velocity ($v_F$)
  versus the angle $\theta_\bk$ with $L=90$nm.
  Dotdashed red, solid green and dashed blue lines correspond to $U_0$ being
  10~meV, 30~meV and 50~meV, respectively.
  (b) The inverse effective mass $m_e/m_{xx}^*$ versus $U_{0}$ for the first band above the Dirac point on BZ center.
  Dotdashed red, solid green and dashed blue lines are results  for $L$ being $30$nm, $60$nm and $90$nm, respectively.  }
  \label{fig:rv}
\end{figure}

In the  limit of a perturbatively weak coupling
to the HSDW ($\frac{|U(\bG)|}{\hbar v_F|\bG|}\ll1$), the  leading order contribution of the HSDW to the low-momentum $(\bm k\sim 0)$
dispersion can be characterized by the renormalization of the group velocity and effective mass at $\bm k=0$.
 The group velocity of states without the HSDW is isotropic around
 ${\bf k}=0$ with constant
magnitude $v_F$. In presence of the HSDW, the renormalized group
 velocity of quasiparticles
parallel to the wavevector $\bk$ [$v_{\hat k}\equiv\bv(\bk)\cdot\hat k$]
around the Dirac point ($|k| \ll 1$) obtained within second order
 perturbation theory can be written as
\begin{equation}
\begin{array}{l l l l l l }
\dfrac{v_{\hat k}-v_F}{v_F}= \dfrac{1}{\hbar v_F}\dfrac{\partial(E_{+,k}-\varepsilon_{+,k})}{\partial k}=
-\dfrac{U_0^2 L^2}{8 \pi^2 \hbar^2 v_F^2}(3+\cos  2\theta_{\bk})
\label{eq:vel}
\end{array}
\end{equation}
From Eq.~(\ref{eq:vel}), it is clear that the renormalized group velocity is anisotropic around the Dirac point and decays monotonically, both with the amplitude of periodic potential $U_0$, and with the spatial period $L$ of the potential. Fig.~\ref{fig:rv} shows the result from a full numerical calculation. We note that there is good agreement between the trends from Eq.~(\ref{eq:vel}) and the full calculation (shown in Fig.~\ref{fig:rv}a) for weak potential strength.

The interaction with the HSDW
introduces at $\bm k\sim0$ a finite curvature along the $\hat{x}$-direction to the previously linear in $\bm k$ dispersion of the Dirac fermions.
The effective mass tensor at ${\bf k}=0$ from the curvature of the energy band around the Dirac point is given by
\begin{equation}
\Big(\dfrac{m_e}{m^*}\Big)_{\alpha\beta} \ = \ \dfrac{m_e}{\hbar^2} \dfrac{\partial^2 E(\bk)}{\partial k_\alpha \partial k_\beta} \Big|_{{\bf k}=0}= \Bigg\{\begin{array}{l l}
0,  \ \ \alpha,\beta=y,\ \text{or}\ \alpha \neq \beta
\\
\\
\dfrac{8 U_0^2 m_e}{\hbar^3 v_F G_0^3}, \ \  \alpha,\beta=x
\label{mass}
\end{array}
\end{equation}
where $m_e$ is the bare electron mass and $G_0 =\dfrac{2\pi}{L}$. From Eq.~(\ref{mass}) the effective mass $m_e/m_{xx}^*$ grows as $U_0^2$, which
describes the small $U_0$ behavior of the numerically determined effective mass shown in Fig.~\ref{fig:rv}b.

\subsection{Dirac Cone near BZ boundary}
\begin{figure}
  \includegraphics[width=1.0\columnwidth]
  {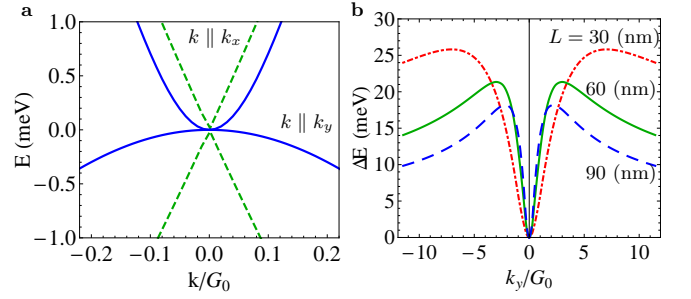}
  \caption{ (Color online)
  (a)
Semi-Dirac charge carrier dispersion versus the wavevector $k$ with $k$ measured from $(k_x, k_y)=(\dfrac{\pi}{L},0)$ on the BZB.
  Dashed (green) and solid (blue) lines show the linear
dispersion along the $\hat{x}$ direction and the quadratic dispersion
 along the $\hat{y}$
direction, respectively with $U_0=50$meV, $L=90$nm and $G_0=2\pi/90$nm$^{-1}$ ($E(\pi/L,0)$ was shifted to zero).
  (b).The energy gap $\Delta E$ between
  the first and the second band at the BZB
  versus $k_y$ for charge carriers above  the Dirac point with $U_0=50$meV and $G_0=2\pi/90$nm$^{-1}$.
  Dotdashed (red) ,  solid(green) and dashed (blue) lines
 correspond to $L$ being
  30~nm, 60~nm and 90~nm, respectively.  }
  \label{fig:delta}
  \end{figure}

While the periodic potential from the HSDW introduces renormalizations
of the velocity and effective masses at the center of the Dirac cone
at $\bm k=0$, the effect of the periodic potential is limited to being
perturbative since there is no degeneracy in the initial Dirac spectrum
near $\bm k=0$. However, such degeneracies do occur at the edges
of the FBZ, which we have referred to as the BZB, and non-perturbative
effects of the potential may be generated by coherent back scattering
with momenta $\bm G$.
Unlike the case of conventional crystals, where such back-scattering
opens up gaps, one finds the emergence of new Dirac cones at the BZB
in the presence of interaction with the HSDW (Fig.~\ref{fig:delta}a).

 When the wavevector $\bk$ is on the first BZB
($(k_x, k_y)=(\pi/L,k_y)$), the two states $\left|s,\bk\right>$ and
$\left|s,\bk-(2\pi/L,0)\right>$ are degenerate before the periodic
 potential is applied. In the presence of an applied potential, the
 largest contribution to the energy eigenvalues at the edges of the
 BZ comes from these two degenerate states. For the clockwise helix, the backscattering amplitude leads to an
 energy gap on the BZB that is given by,
\begin{equation}
\Delta E(k_y) = U_0\left(1-\frac{s}{\sqrt{1+(k_y L/\pi)^2}}\right),
\label{eq:gap1D}
\end{equation}
where $s=\pm 1$ represents  the positive
or negative bands in the Dirac cone. In the lower band of the Dirac cone,
i.e. $s=-1$, the above gap is maximum at $k_y=0$ and decreases
 monotonically with $|k_y|$. The positive band is more interesting with
a gap that vanishes at $k_y=0$. From the full numerical calculation in Fig. \ref{fig:delta}b, we can see that the energy gap of the positive band on the BZB increases monotonically then decreases with $k_y$.

If we change the sign of both $U_y$ and $U_z$, the physics will not change since the total Hamiltonian is invariant under this transformation $T$, where $T$ is the operator corresponding to translation by
$L/2$ in the $\hat{x}$ direction. The chirality of the helix depends on the relative sign of the $U_y$ and $U_z$ in the HSDW potential. The effects of the clockwise helix on the upper band of the Dirac cone is the same as the anticlockwise helix on the lower band of the Dirac cone. For the anticlockwise helix, the minus sign in Eq. \ref{eq:gap1D} should be changed to a plus sign, which leads to a vanishing gap at $k_y=0$ for the lower band.

  \begin{figure}
  \includegraphics[width=1.0\columnwidth]
  {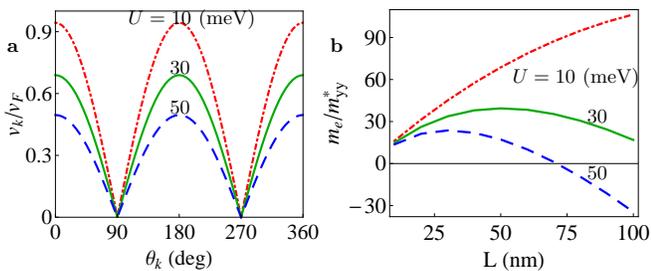}
  \caption{(Color online) \ (a) Anisotropic $v_{\hat k}/v_F$ measured from the
 semi-Dirac point on the BZB
  versus the angle $\theta_\bk$  with $L=90$nm.
  Dotdashed (red), solid (green) and dashed (blue)
 lines correspond to $U_0$ being
  10~meV, 30~meV and 50~meV, respectively. (b)
  The inverse effective mass $m_e/m_{yy}^*$ versus $L$
  around the point $(k_x,k_y)=(\dfrac{\pi}{L},0)$ for the lower
 band of the semi-Dirac point for three $U_0$. Note the
 negative effective mass (corresponding to Fig. \ref{fig:delta} (a)). }
  \label{fig:ZIvemU}
  \end{figure}

As with the Dirac cone at $\bm k=0$, an understanding of the low-energy
dispersion and the corresponding transport properties associated with the
Dirac cone at the BZB is obtained by calculating the group velocity and
effective mass. Second order perturbation theory leads to a strongly
anisotropic group-velocity  $v_{\hat{k}}$ measured from
the gap-less point at $(k_x,k_y)=(\dfrac{\pi}{L},0)$ which is written as
\beq
v_{\hat{k}}=\dfrac{\hbar v_F^2 G_0 \cos\theta_\bk}{\sqrt{U_0^2+\hbar^2 v_F^2 G_0^2}}
\label{eq:ZID}
\eeq
For comparison, the full numerical calculation is shown in Fig.\ref{fig:ZIvemU}. The effective mass tensor at the new gapless point for the clockwise helix, calculated within second order perturbation theory similar to Eq. \ref{mass}, is given by
\begin{equation}
\Big(\dfrac{m_e}{m^*}\Big)_{\alpha\beta}\Big|_{{\bf k}=(\pi/L,0)}=\Bigg\{\begin{array}{l l}
0,  \ \ \ \ \ \alpha, \beta=x,\ \text{or}\ \alpha \neq \beta
\\
\\
2(\hbar v_F G_0\pm U_0)/G_0^2, \ \ \ \alpha, \beta=y
\label{eq:ZIE}
\end{array}
\end{equation}
where the $\pm$  corresponds to the upper or lower band around the semi-Dirac point at $(k_x,k_y)=(\dfrac{\pi}{L},0)$. Eq. \ref{eq:ZID} and \ref{eq:ZIE} together show the highly anisotropic dispersion of the semi-Dirac point on BZB: a dispersion linear along the periodic direction $\hat{x}$ but quadratic along the perpendicular direction $\hat{y}$. For the anticlockwise helix, the effective mass tensor given in Eq. \ref{eq:ZIE} is still valid except now the gap for the lower band vanishes at $k_y=0$. Fig. \ref{fig:ZIvemU}a shows the angular dependence of the group velocity $v_{\bk}$ from a full numerical calculation. The velocity $v_y$ is seen to have zero component regardless of the strength of the applied potential. The dependence of the renormalized group velocity $v_k/v_F$  is a decreasing function of the amplitude of the applied potential. In Fig. \ref{fig:ZIvemU}b, we plot the dependence of the inverse effective mass of the lower-band
of the semi-Dirac point at the BZB
(in the vicinity of $(k_x,k_y)=(\dfrac{\pi}{L},0)$)
 on the applied potential period $L$ and amplitude $U_0$ respectively.
 We find that the effective mass along the $\hat{y}$-direction
changes sign from positive to negative as a function of $L$ and $U_0$.
The dispersion at small $k_y$ for parameters
$U,L_0$ corresponding to negative
effective mass is shown in Fig. 3(a). In this case, the bands
at larger $k_y$ turn up leading to the emergence of another
fermi surface. There is no such fermi surface doubling for parameters
with positive mass.

\section{Other forms of spin density wave}
\label{sec:other}
\subsection{The HSDW with $U_z\neq U_y$}
\label{sec:yz}
\begin{figure}\includegraphics[width=1.0\columnwidth]{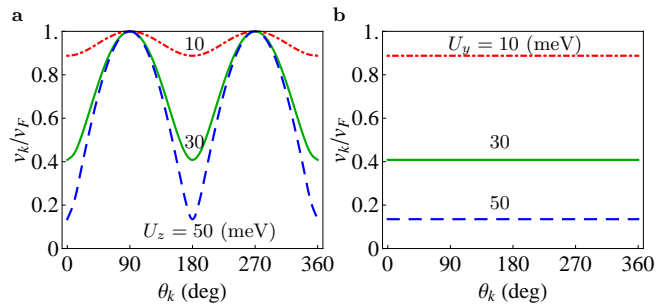}
  \caption{(Color online)
  (a) The group velocity $v_{\hat{k}}$ (measured from the Dirac point)
  of charge carriers on the surface of the TI in units of the Fermi velocity ($v_F$)
  versus the angle $\theta_\bk$ with $L=90$nm and $U_y=0$.
  Dotdashed red, solid green and dashed blue lines correspond to $U_z$ being
  10~meV, 30~meV and 50~meV, respectively.
  (b) The group velocity $v_{\hat{k}}$ (measured from the Dirac point)
  of charge carriers on the surface of the TI in units of the Fermi velocity ($v_F$)
  versus the angle $\theta_\bk$ with $L=90$nm and $U_z=0$.
  Dotdashed red, solid green and dashed blue lines correspond to $U_y$ being
  10~meV, 30~meV and 50~meV, respectively.}
  \label{fig:rvyz}
\end{figure}

In this subsection, we shall discuss the case of a HSDW with different amplitudes in the $\hat{y}$ and $\hat{z}$ directions (i.e. $U_y \neq U_z$) with the propagation direction of the spin density wave still along the $\hat{x}$ direction. In order to study the effects of helical spin density wave with $U_z \neq U_y$, we first use the perturbation theory to calculate the renormalized velocity in the BZ center. In the  limit of a perturbatively weak coupling
to the HSDW ($\frac{|U(\bG)|}{\hbar v_F|\bG|}\ll1$), the renormalization of group velocity (near the Dirac cone) given in Eq. \ref{eq:vel} within second order
 perturbation theory should be modified to:
\begin{equation}
\begin{array}{l l l l l l }
\dfrac{v_{\hat k}-v_F}{v_F}= -\dfrac{L^2}{8 \pi^2 \hbar^2 v_F^2}\big[2 U_y^2+U_z^2 (1+\cos  2\theta_{\bk})\big]
\label{eq:velyz}
\end{array}
\end{equation}
It is clear that the above formula goes to Eq. \ref{eq:vel} for the proper HSDW. We notice that the spin density wave in the $\hat{y}$ direction does not induce the anisotropic group velocity near the Dirac cone, which is also confirmed by the full numerical calculation. We present the angular dependence of the renormalized group velocity for two extreme limits $U_y=0$ and $U_z=0$ in Fig. \ref{fig:rvyz}.

\begin{figure}
  \includegraphics[width=1.0\columnwidth]
  {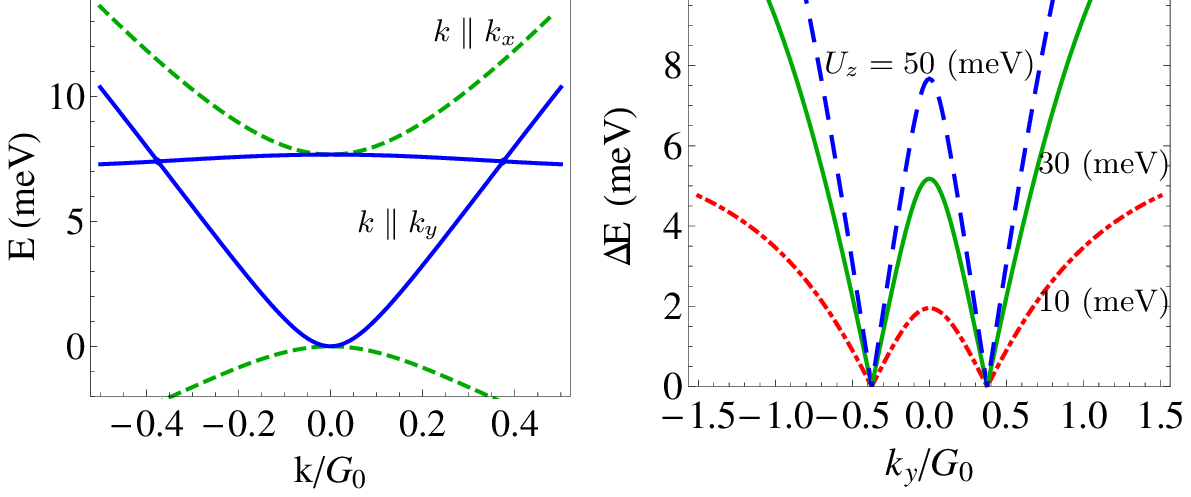}
  \caption{(Color online) \ (a) Energy charge carrier dispersion versus the wavevector $k$ with $k$ measured from $(k_x, k_y)=(\dfrac{\pi}{L},0)$ on the BZB.
  Dashed (green) and solid (blue) lines show the linear
dispersion along the $\hat{x}$ direction and the quadratic dispersion
 along the $\hat{y}$
direction, respectively with $U_z=50$meV, $U_y=40$meV,  $L=90$nm and $G_0=2\pi/90$nm$^{-1}$  ($E(\pi/L,0)$ was shifted to zero).  (b)The energy gap $\Delta E$ between
  the first and the second band at the BZB
  versus $k_y$ for charge carriers above  the Dirac point with $L=90$nm and $U_y=0.8U_z$.
  Dotdashed (red) ,  solid(green) and dashed (blue) lines
 correspond to $U_z$ being
  10~meV, 30~meV and 50~meV, respectively. }
  \label{fig:Uzy}
 \end{figure}

\begin{figure}
  \includegraphics[width=1.0\columnwidth]
  {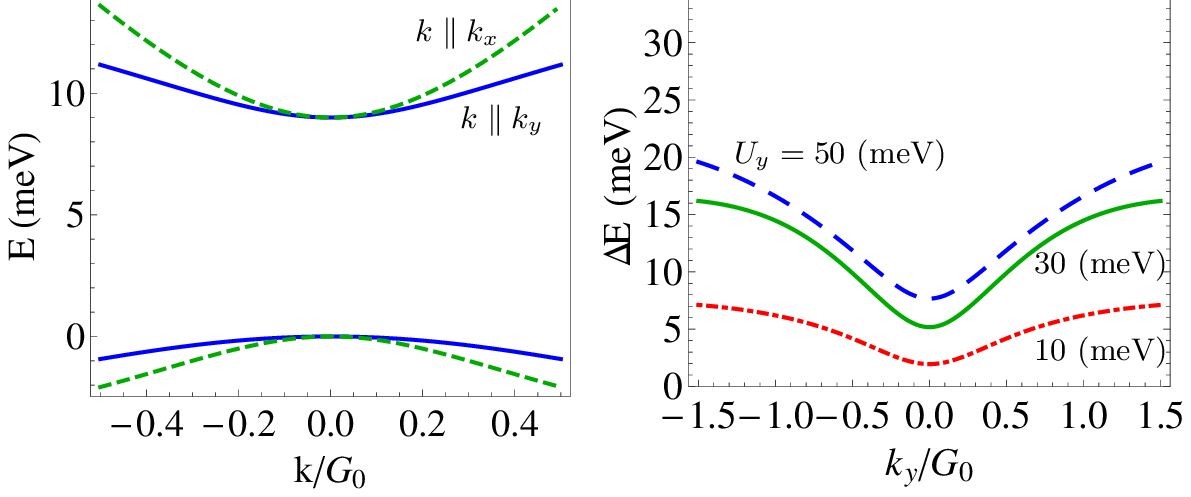}
  \caption{(Color online) \ (a)
  Energy charge carrier dispersion versus the wavevector $k$ with $k$ measured from $(k_x, k_y)=(\dfrac{\pi}{L},0)$ on the BZB.
  Dashed (green) and solid (blue) lines show the linear
dispersion along the $\hat{x}$ direction and the quadratic dispersion
 along the $\hat{y}$ direction, respectively with $U_z=40$meV, $U_y=50$meV,  $L=90$nm and $G_0=2\pi/90$nm$^{-1}$  ($E(\pi/L,0)$ was shifted to zero). (b)The energy gap $\Delta E$ between
  the first and the second band at the BZB
  versus $k_y$ for charge carriers above  the Dirac point with $L=90$nm and $U_z=0.8U_y$.
  Dotdashed (red) ,  solid(green) and dashed (blue) lines
 correspond to $U_y$ being
  10~meV, 30~meV and 50~meV, respectively. }
  \label{fig:Uyz}
 \end{figure}

\begin{figure}
  \includegraphics[width=1.0\columnwidth]
  {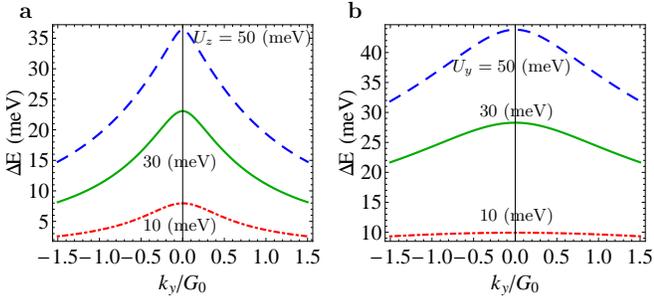}
  \caption{(Color online) \ (a)
The energy gap $\Delta E$ between
  the first and the second band at the BZB
  versus $k_y$ for charge carriers above  the Dirac point with $L=90$nm and $U_y=0$.
  Dotdashed (red) ,  solid(green) and dashed (blue) lines
 correspond to $U_z$ being
  10~meV, 30~meV and 50~meV, respectively. (b)The energy gap $\Delta E$ between
  the first and the second band at the BZB
  versus $k_y$ for charge carriers above  the Dirac point with $L=90$nm and $U_z=0$.
  Dotdashed (red) ,  solid(green) and dashed (blue) lines
 correspond to $U_y$ being
  10~meV, 30~meV and 50~meV, respectively. }
  \label{fig:onlyone}
 \end{figure}

We shall next explain the conditions for the appearance of the Dirac cones on the BZ boundary, which are supported by full numerical calculations as presented in Fig. \ref{fig:Uzy} and \ref{fig:Uyz}. The largest contribution to the energy eigenvalues at the BZ boundary
comes from these two degenerate states $\left|s,\bk\right>$ and
$\left|s,\bk-(2\pi/L,0)\right>$, which  are degenerate before applying the
periodic potential. Only considering the above two states and using the relation $\theta_{\bk-(2\pi/L,0)}=\pi-\theta_{\bk}$ on the BZB, we can write down the two band Hamiltonian within perturbation theory as:
\begin{widetext}
\begin{equation}
H(\bk) = \left(
\begin{array}{cc}
s\hbar v_F \sqrt{\big(\dfrac{\pi}{L}\big)^2+k_y^2} & \frac{1}{2} (sU_y-U_z \cos\theta_\bk) (i \cos\theta_\bk+\sin\theta_\bk)\\
\frac{1}{2} (sU_y-U_z \cos\theta_\bk) (-i \cos\theta_\bk+\sin\theta_\bk)& s\hbar v_F \sqrt{\big(\dfrac{\pi}{L}\big)^2+k_y^2}
\end{array}
\right)\ ,
\label{eq:2band}
\end{equation}
\end{widetext}
where $s=\pm1$ denote the band index and $\theta_\bk$ is the angle between the vector $\bk$ and the $\hat{x}$ direction.
Diagonalizing the above Hamiltonian we could get two eigenvalues written as:
\begin{equation}
\begin{array}{l l l l l }
\epsilon_1= s\hbar v_F \sqrt{\big(\dfrac{\pi}{L}\big)^2+k_y^2} -\dfrac{1}{2}\sqrt{(sU_y-U_z \cos\theta_\bk)^2}
\\
\\
\epsilon_2=s\hbar v_F \sqrt{\big(\dfrac{\pi}{L}\big)^2+k_y^2} +\dfrac{1}{2}\sqrt{(sU_y-U_z \cos\theta_\bk)^2}
\\
\\
\Delta E = \mid sU_y-U_z \cos\theta_\bk \mid,
\end{array}
\label{eq:deltayz}
\end{equation}
From the last equation in Eq. \ref{eq:deltayz}, we can clearly see that there are gapless states when $sU_y=U_z \cos\theta_\bk$, since $0 < \cos\theta_\bk \leq 1$ for $\bk$ on the BZ boundary (i.e. $\bk = (\pi/L,k_y)$). Therefore, we can have gapless states only for $U_z \geq U_y$. In addition, the energy charge carrier dispersion becomes parabolic along both $\hat{x}$ and $\hat{y}$ directions, which can be seen from Fig. \ref{fig:Uzy}(a) and \ref{fig:Uyz}(a). For the clockwise helix, the energy gap opened on the BZ boundary near $\bk = (\pi/L, 0)$ is shown as a function of $k_y$ for three different values of $U_{z,y}$. In Fig. \ref{fig:Uzy}(b) and Fig. \ref{fig:Uyz}(b) the energy gap is plotted as a function of $k_y$ for $U_y/U_z=0.8$ and $U_z/U_y=0.8$ respectively. It is seen that the energy gap is non-monotonic function of $|k_y|$ in the former, and monotonic in the latter case. The helical spin density with $U_y \neq U_z$ affect the energy charge carrier dispersion on the BZ center similar to the proper HSDW, rendering the Dirac cone on the BZ center highly anisotropic. The velocity of the Dirac cone along $\hat{x}$ direction (the helix propagates along the $\hat{x}$ direction) is smaller than that perpendicular to $\hat{x}$. When we go to the extreme limit , where one of the components $U_{y,z}$ is set to zero (the spin density wave), the energy gap in the presence of the spin density wave with magnetization in only one direction ($\hat{y}$ or $\hat{z}$), further simplifies to:
\begin{equation}
\Delta E = \Bigg\{
\begin{array}{l l l l l }
\mid U_z \cos\theta_\bk \mid, \ \ \ \ U_y=0
\\
\\
\mid U_y \mid, \ \ \ \ \ \ \ \ \ \ \ \ U_z=0
\end{array}
\label{uyuz}
\end{equation}
Full numerical calculation shows (see Fig \ref{fig:onlyone}) the dependence on $k_y$ for the energy gap opened on the BZB with the presence of the spin density waves in either $\hat{z}$ or $\hat{y}$ direction. The latter dependence is not predicted correctly within the second order perturbation theory which suggests that the energy gap for $U_z=0$ is independent of $k_y$ (see second line of Eq. \ref{uyuz}). Finally, one should note that the Hamiltonian has different symmetries along $\hat{y}$ and $\hat{z}$ directions. For a spin density wave only in the $\hat{y}$ direction, the total Hamiltonian satisfies the relation: $\sigma_z H \sigma_z= -H$, from which we can see that the upper and the lower band of the Dirac cone are symmetric. While for the spin density wave only in the $\hat{z}$ direction, the total Hamiltonian satisfies particle-hole symmetry, i.e. $\sigma_x H^*(-\bk) \sigma_x= -H(\bk)$. The upper and the lower band of the Dirac cone are also symmetric when the spin density wave only has component in $\hat{z}$ direction. The velocity perpendicular to the $\hat{x}$, which is the same as the original Fermi velocity of $Bi_2Se_3$, remains unaffected in the presence of the periodic potential in the $\hat{z}$ direction.

\subsection{Cycloidal spin density wave}
In this subsection, we analyze the effects of the cycloidal spin density wave on the surface states of topological insulator. In the presence of the cycloidal spin density wave on top of the topological insulator\cite{bogdanov_JPCS_2010,sosnowska_JPSJ_2006}, the potential is modified to:
\begin{equation}
\begin{array}{l l }
U_c(x)= U_x \sigma_x \cos(\dfrac{2\pi}{L}x)+ U_{y,z} \sigma_{y,z} \sin(\dfrac{2\pi}{L}x),
\end{array}
\end{equation}
The term involving $\sigma_x$ in the magnetic potential will disappear under the gauge transformation:
\begin{equation}
\begin{array}{l l }
e^{-i A(x)} H e^{i A(x)}= H_0+ U_{y,z} \sigma_{y,z} \sin(\dfrac{2\pi}{L}x),
\end{array}
\end{equation}
where $H_0$ is the single Dirac fermion Hamiltonian without applied potential as given in Eq. \ref{ham} and the function $A(x)$ is given by:
\beq
A(x)=-\dfrac{U_x L}{ 2\pi \hbar}\sin\big(\dfrac{2\pi x}{L}\big)
\eeq
For the general periodic function of spin density wave along $\hat{x}$ direction, i.e. $F(x) \sigma_x$, the $\sigma_x$ term will disappear if we apply the gauge transformation $e^{i A_F(x)}$ with $A_F(x)=-\int^x_{c} F(x')/\hbar dx'$ ($c$ is any constant). Thus, the effects of the cycloidal spin density wave comes from the term with $\sigma_{y,z}$, which goes to the extreme limit as discussed in Sec.\ref{sec:yz}.

\section{Summary and Conclusions}
\label{sec:summary}
In summary, we have considered the effects of a helical spin density wave on the surface states of a topological insulator and also the effects of other spin density wave on the surface transport property of topological insulator. We find that the HSDW acts like an
 effective spin potential on the TI surface and breaks the local time reversal invariance in the latter. The applied spin potential has two main consequences:
 First, the group velocity of electrons at the Dirac point is strongly suppressed in the direction transverse to the periodic potential. Secondly, new semi-Dirac
 points emerge for the first upper (lower) band at the Brillouin zone boundaries that corresponds to the right-handed (left-handed) chirality of the applied proper HSDW. The semi-Dirac points are characterized by linear dispersion parallel to the direction $\hat{x}$ of the applied periodic potential but quadratic dispersion perpendicular to the $\hat{x}$ direction. When the chemical potential of the TI is such that the semi-Dirac points at the BZBs give the main contribution to
transport,
we expect the  surface transport properties to be highly anisotropic. This is because at these points, the dispersion is linear in one
direction and quadratic in the other. The momentum space location of these new semi-Dirac cones can also
 be manipulated by applying an in-plane magnetic field to the HSDW. Such a magnetic field \cite{cheong_nature_2007} can change the
 orientation of the spin rotation axis relative to  the pitch vector of the HSDW. By studying the effects of such a change in the applied periodic potential on the new semi-Dirac cones and measuring the anisotropic component of the transport, it should be possible to isolate the surface state contribution  from the bulk
 in the total conductance in topological insulators with relatively small bulk gaps. The number of Dirac cones appearing on the BZ boundary is determined by the relative ratio of $U_z/U_y$, which could be zero, one or two corresponding to $U_z/U_y<1$, $U_z/U_y=1$ and $U_z/U_y>1$, respectively. We also prove that the spin density wave with spin vectors along $\hat{x}$ direction will not change the surface transport property of a three dimensional topological insulator. One particular advantage of our proposal is that the proposed experiment can be carried out in the existing TI systems without any problem arising from the bulk conduction (as a result of unintentional bulk doping by defects), which is invariably present in almost all current TI systems, since the bulk transport is isotropic and presumably unaffected by the presence of the HSDW.

Although the proposed system in this work is a sandwich structure of a TI and a multiferroic, we believe that this structure may turn out to be a suitable candidate for the direct manifestation of the elusive 2D transport properties even in the presence of considerable bulk conduction since the anisotropy introduced by the presence of the HSDW would only affect the surface transport properties without affecting much the bulk conduction behavior.  Given the great current interest and activity in the observation of the 2D surface transport properties in TI materials \cite{Ren_PRB10, Qu_Science10, Butch_PRB10, Analytis_nph_2010, Culcer_PRB10, Checkelsky_arXiv10, Taskin_PRB10, Akinori_PRB10, Eto_PRB10, DasSarma_arXiv10}, we are optimistic that our proposed structure could go a long way in establishing the experimental behavior of 2D surface transport in topological insulators.

\begin{acknowledgments}
 Q.L. acknowledges helpful discussions with Kai Sun. Q.L., J.D.S. and S.D.S. are supported by DARPA-QuEST, JQI-NSF-PFC. P.G. is supported by National Institute of Standards and Technology through Grant Number 70NANB7H6138, Am 001 and through Grant Number N000-14-09-1-1025A by the Office of Naval Research. S.T. acknowledges support from AFOSR and Clemson University start up funds.
\end{acknowledgments}

\end{document}